\documentclass[aps,prstab,reprint,groupedaddress,showpacs,amsmath,amssymb]{revtex4-1}
\usepackage{amssymb}
\usepackage{amsfonts}
\usepackage{epsfig}
\usepackage{dcolumn}
\usepackage{bm}
\usepackage{amsmath}
\usepackage{graphicx}
\usepackage{subfigure}

\begin{document}

\title{Field Stabilization of Alvarez-Type Cavities}
l
\author{X.~Du, L.~Groening, S. Mickat, A. Seibel}
\affiliation{GSI Helmholtzzentrum f\"ur Schwerionenforschung GmbH, D-64291 Darmstadt, Germany}


\date{\today}

\begin{abstract}
Alvarez-type cavities are commonly used to reliably accelerate high quality hadron beams. Optimization of their longitudinal field homogeneity is usually accomplished by post-couplers, i.e. additional rods being integrated into the cavity. This paper instead proposes to use the stems that keep the drift tubes for that purpose. As their individual azimuthal orientations do not change the cavity's undisturbed operational mode, they comprise a set of free parameters that can be used to modify higher mode field patterns. The latter have significant impact on the robustness of the operational mode w.r.t. eventual perturbations. Several optimized stem configurations are presented and benchmarked against each other. The path to obtain these configurations is paved analytically and worked out in detail through simulations. It is shown that the method provides for flat field distributions and very low field tilt sensitivities without insertion of post-couplers.
\end{abstract}

\maketitle


\section{Introduction}
In 1945 L.~Alvarez invented rf-cavities to accelerate beams of protons~\cite{AI}. These cavities became the work horses of many drift tube linacs~(DTLs) all over the world~\cite{Linac4,HIDIF DTL}. They provide a longitudinally constant axial electric and an azimuthally constant magnetic field being referred to as the $TM_{010}$ operating mode. Along each cell the radio-frequency (rf)-fields have the same phase implying that the phase advance between adjacent cells is zero. The group velocity $v_g$ of a resonating mode is proportional to the slope $\delta\omega/\delta k$ of the dispersion curve. Since the group velocity of the $TM_{010}$ mode is zero, there is no power flux along the structure and the field distribution is very sensitive to frequency perturbations of the cells, to power losses, and to beam loading. This problem is particularly serious for long DTLs, where the frequency separation between modes is low~\cite{Wangler1}. Frequency perturbations usually arise from mechanical errors of the cavity and from thermal distortion during operation which cannot be mitigated completely. For these reasons field stabilization is a major concern in the design and operation of Alvarez-type cavities.

The common solution for field stabilization has been proposed in the 1960ies at the Los Alamos Notional Laboratory~\cite{PC1960}. Internal metallic rods extend from the cavity mantle towards the drift tube~(DT) without touching them. These post-couplers (PCs) form a resonant structure that couples to the operational mode of the cavity~\cite{PCtuning}. However, this method does not apply to large tank radii and previous experience with field stabilization of large diameter Alvarez-type cavities shows that stabilization with standard PCs is very difficult or even impossible~\cite{MV,twostem}.

Alvarez-type cavities with diameters exceeding about one meter need to provide two stems per tube for mechanical robustness. For such cases this paper proposes to use the stems itself for field tuning. A specific tuning method is developed based on dedicated stem configurations to provide for the function of post-couplers in large diameter cavities.

In the following section an equivalent circuit for a cavity equipped with stems is introduced together with 3d-field simulations for an Alvarez-type cavity with constant particle velocity profile. The third section is on dispersion relations from the equivalent circuit and presents the principle of stabilization through orientation of the stems. Afterwards different tuning strategies are established to achieve field flatness optimization and low field tilt sensitivities by simulations during the cavity design stage. In the fifth section perturbations are added to analyse the achievable field tilt sensitivity~(TS). The efficiency of the proposed tuning methods is demonstrated through simulations of a cavity with increasing cell length.

\section{Equivalent Circuit of Alvarez-Type Cavity }
Previous studies showed that the stems play a significant role in the shaping of dispersion curves of $TM$-modes~\cite{LA SG}. During simulations for the UNILAC post-stripper upgrade~\cite{UNILAC_Upgrade} a considerable influence of the stem configuration on the mode spectrum and on the TS~\cite{Xdu} was observed. The equivalent circuit is studied in order to benefit from this effect by introducing a new field stabilization method.

\begin{figure}[hbt]
\centering
\includegraphics*[width=80mm,clip=]{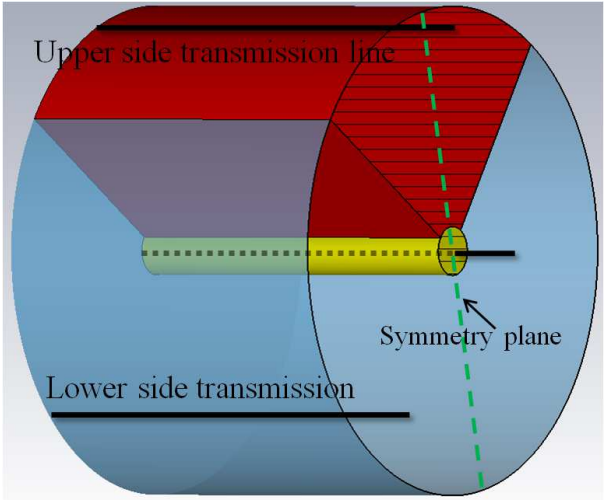}
\caption{Virtual division of an Alvarez-type cavity into three sections to be modelled by one transmission line each.}
\label{Divide}
\end{figure}

\begin{figure}[hbt]
\centering
\includegraphics*[width=80mm,clip=]{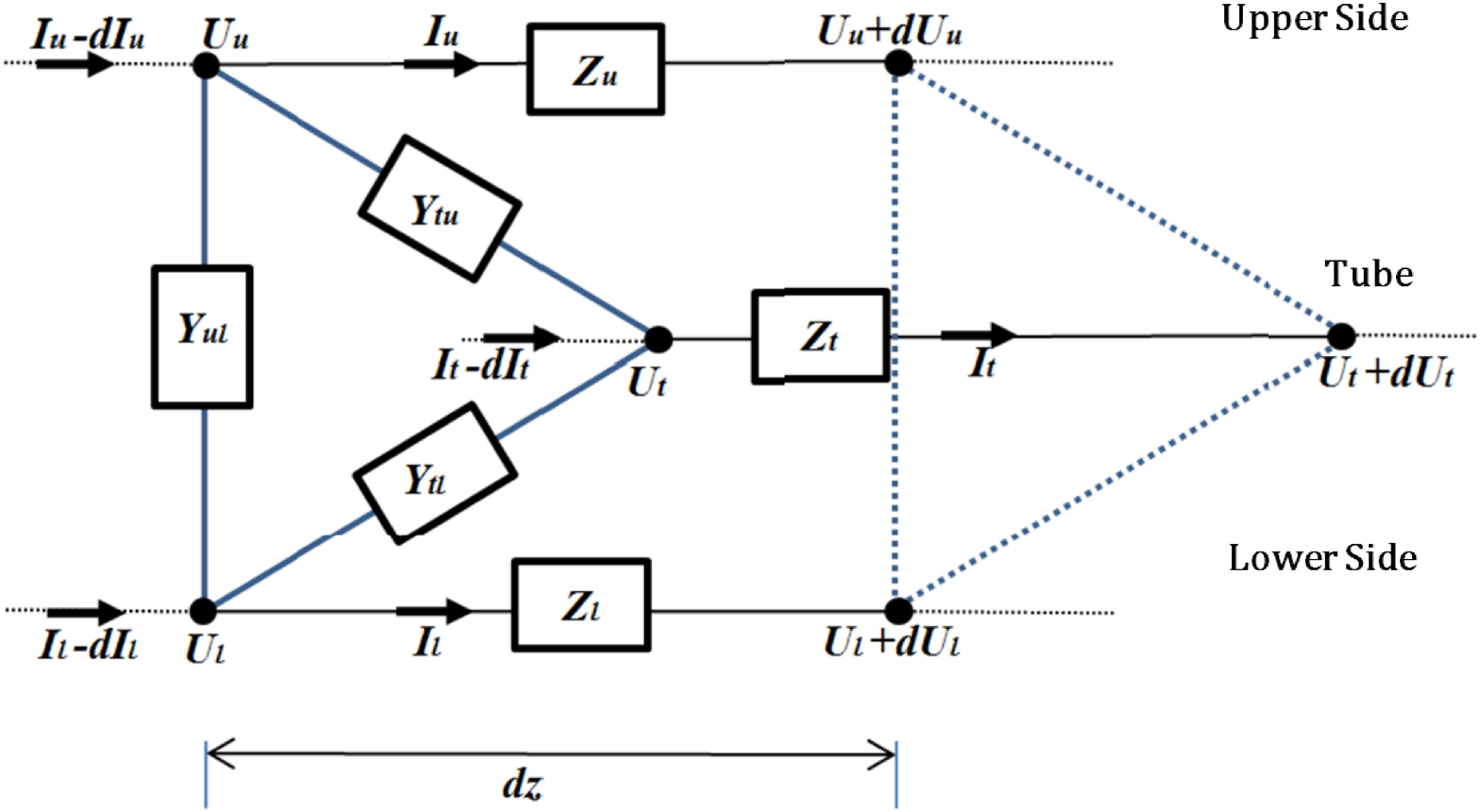}
\caption{Equivalent circuit from three transmission lines for an Alvarez-type cavity.}
\label{3T}
\end{figure}

In general, operational Alvarez-type cavities have one or two sets of uniform stems keeping the drift tubes on the axis of the cavity. It is convenient to virtually divide the cavity in an upper part, a lower part, and the drift tubes~(Fig.~\ref{Divide}). This allows for introduction of three conductor transmission lines being assigned along the tubes, the upper, and the lower part of the tank. This topology is similar to the four conductor line introduced in~\cite{Transmision line}. The analogue analysis of $\pi-$mode structures without stems has been developed and applied to field tuning in~\cite{Pi_structure,Schriber} but tuning with PCs or stems requires more specific analysis.
The resulting transmission line model along a length $dz$ on the beam axis being short w.r.t. the tank length is shown in Fig.~\ref{3T}. In this circuit $Z_u$, $Z_l$, and $Z_t$ are the longitudinal impedances per length on the upper and lower side of the tank and of the tube array. $Y_{tu}$, $Y_{tl}$, and $Y_{ul}$  are the transverse admittances between tube and upper side, tube and lower side, and between upper side and lower side. The parameters are indexed referring to the three parted DTL cross-section introduced above: $u$ (upper part), $l$ (lower part), and $t$ (tube). In the same way double indices are used, e.g. $Y_{tu}$ as transverse admittance between the tube and the upper part of the DTL per length, $Y_{tl}$ as transverse admittance between the tube and lower part of the DTL, and $Y_{ul}$ as transverse admittance between upper and lower part of the DTL. These admittances per length are independent of the length $dz$. In the further analysis $dz$ will be defined as one cell length $\beta\lambda$ comprising two half tubes and one gap implying the assumption that the overall tank length is large compared to the single cell length.

\subsection{Differential Equations of Field Distribution}
Applying the sum rule to the equivalent circuit of Fig.~\ref{3T}, the differential voltages along the cavity are given by the longitudinal current through
\begin{equation}
    \begin{aligned}
d U_t=Z_t I_t\\
d U_u=Z_u I_u\\
d U_l=Z_l I_l\,,\\
    \end{aligned}
\end{equation}
where the change of longitudinal current is the sum of the transverse currents
\begin{equation}
    \begin{aligned}
d I_t=Y_{tu}(U_u-U_t)+Y_{tl}(U_l-U_t)\\
d I_u=Y_{tu}(U_t-U_u)+Y_{ul}(U_l-U_u)\\
d I_l=Y_{tl}(U_t-U_l)+Y_{ul}(U_u-U_l)\,.\\
    \end{aligned}
\end{equation}
These mesh equations deliver quadratic differential equations for the voltage distribution along the equivalent transmission line
\begin{equation}
\begin{bmatrix}
\begin{smallmatrix}
\frac{d^2U_t}{dz^2}\\
\frac{d^2U_{u}}{dz^2}\\
\frac{d^2U_{l}}{dz^2}
\end{smallmatrix}
\end{bmatrix}
=
A
\begin{bmatrix}
\begin{smallmatrix}
U_t\\
U_u\\
U_l
\end{smallmatrix}
\end{bmatrix}\,,
\end{equation}
with the coefficient matrix
\begin{equation}
A
:=
\begin{bmatrix}
\begin{smallmatrix}
Z_t (Y_{tu}+Y_{tl})&  -Z_t Y_{tu} &  -Z_t Y_{tl}\\
-Z_u Y_{tu} &  Z_u (Y_{tu}+Y_{ul}) &  Z_uY_{ul} \\
-Z_l Y_{tl} &  Z_l Y_{ul} &  Z_l(Y_{tu}+Y_{ul})
\end{smallmatrix}
\end{bmatrix}\,.
\end{equation}
Defining
$$
H
=
\begin{bmatrix}
\begin{smallmatrix}
1&  0 &  0\\
1&  -1 &  0 \\
1&  0 &  -1
\end{smallmatrix}
\end{bmatrix}
$$
with the transformation
\begin{equation}
H
\begin{bmatrix}
\begin{smallmatrix}
\frac{d^2U_t}{dz^2}\\
\frac{d^2U_{u}}{dz^2}\\
\frac{d^2U_{l}}{dz^2}
\end{smallmatrix}
\end{bmatrix}
=
HA H^{-1} H
\begin{bmatrix}
\begin{smallmatrix}
U_t\\
U_u\\
U_l
\end{smallmatrix}
\end{bmatrix}
\end{equation}
and after some substitutions and simplifications one obtains that the system is described by
\begin{equation}
\label{ddu}
\begin{bmatrix}
\begin{smallmatrix}
\frac{d^2U_t}{dz^2}\\
\frac{d^2U_{tu}}{dz^2}\\
\frac{d^2U_{tl}}{dz^2}
\end{smallmatrix}
\end{bmatrix}
=
B
\begin{bmatrix}
\begin{smallmatrix}
U_t\\
U_{tu}\\
U_{tl}
\end{smallmatrix}
\end{bmatrix}\,,
\end{equation}
where $U_t$ is the voltage along the tube, $U_{tu}=U_t-U_u$ is the voltage between the drift tube and the upper side of the tank, and $U_{tl}=U_t-U_l$ is the voltage between the drift tube and the lower side of the tank. The coefficient matrix $B$ is defined as
\begin{equation}
B=HA H^{-1}=
\begin{bmatrix}
\begin{smallmatrix}
0&  Z_t Y_{tu} &  Z_t Y_{tl}\\
0 &  Z_t Y_{tu}+Z_u (Y_{tu}+Y_{ul}) &  Z_t Y_{tl}-Z_u Y_{ul} \\
0 &  Z_t Y_{tu}-Z_l Y_{ul} &  Z_t Y_{tl}+Z_l (Y_{tl}+Y_{ul})
\end{smallmatrix}
\end{bmatrix}\,.
\end{equation}
These quadratic differential equations can be solved if the distributions of $Z(z)$ and $Y(z)$ are known. Generally $Z(z)$ and $Y(z)$ are designed to be constant to achieve a constant electric field along the cavity axis. This allows for analytical solutions of the voltage distributions, which deliver different field patterns for different modes.

\subsection{Transverse admittances}
Equation~(\ref{ddu}) can be used to derive the tilt of the electric field $E_z$. This tilt should be kept small for the operation of real cavities which are subject to geometrical imperfections
\begin{equation}
\label{Etilt}
\frac{d^2U_t}{dz^2}=\frac{dE_z}{dz}=Z_t(Y_{tu}U_{tu}+Y_{tl}U_{tl})\,.
\end{equation}
Since $Z_t$ is defined by the geometry of the tubes and the tank itself, it must be constant for a given design. The key to minimize the field tilt is minimizing the transverse admittances $Y_{tu}$ and $Y_{tl}$. The admittances can be varied by the angle between subsequent pairs of stems, the stems density on the upper and lower sides of the tank, the number of stems, and by the diameters of the stems.
PCs serve also do adjust the transverse admittance by properly choosing their length~\cite{FG}.
In order to investigate how exactly the stems configuration affects transverse admittances and mode spectra, systematic studies on different stem configurations were conducted. It may be hard to measure or to calculate the values of each inductance and capacitance, not to mention the coupling and mutual inductances among them. A feasible way to determine the values is by deriving the dispersion relations as functions of these unknown values and to conclude these values from few simulated frequencies of $TE$- and $TM$-modes.

\section{Dispersion Relation for different stem configurations}
In order to simplify the analytical treatment but without losing generality, $\beta\lambda$, impedances, and admittances are assumed to be constant along the beam axis. Using the boundary condition that on the front wall~(z=0) and on the end wall~(z=$l$) with~$l$ as the length of the cavity
\begin{equation}
\frac{d^2U_t}{dz^2}=\frac{d^2U_{tu}}{dz^2}=\frac{d^2U_{tl}}{dz^2}=0\,,
\end{equation}
the quadratic differential equations Equ.~(\ref{ddu}) are solved by a standing wave. Accordingly, the voltage distribution for mode number $n$ can be expressed as
\begin{equation}
\label{10}
\begin{bmatrix}
U_t\\
U_{tu}\\
U_{tl}
\end{bmatrix}
=
cos(\frac{n\pi z}{l})
\begin{bmatrix}
U_{t0}\\
U_{tu0}\\
U_{tl0}
\end{bmatrix}\,,
\end{equation}
where $U_{t0},\,U_{tu0}$, and $U_{tl0}$ are constant and
\begin{equation}
\label{11}
\begin{bmatrix}
\frac{d^2U_t}{dz^2}\\
\frac{d^2U_{tu}}{dz^2}\\
\frac{d^2U_{tl}}{dz^2}
\end{bmatrix}
=
B
\begin{bmatrix}
U_{t}\\
U_{tu}\\
U_{tl}
\end{bmatrix}
=
-\frac{n^2\pi^2}{l^2}
\begin{bmatrix}
U_{t}\\
U_{tu}\\
U_{tl}
\end{bmatrix}\,.
\end{equation}
Equation~(\ref{11}) delivers $-\frac{n^2\pi^2}{l^2}$ as an eigen value for $B$ satisfying
\begin{equation}
det(B+\frac{n^2\pi^2}{l^2}I)=0\,,
\end{equation}
where $I$ is the unit matrix. With some transformations the dispersion relation is derived as
\begin{equation}
\label{disp}
\begin{aligned}
(Z_t Y_{tu}+Z_u(Y_{tu}+Y_{ul})+\frac{n^2\pi^2}{l^2})(Z_t Y_{tl}+Z_l(Y_{tl}+Y_{ul})\\
+\frac{n^2\pi^2}{l^2})-(Z_t Y_{tu}-Z_l Y_{ul})(Z_t Y_{tl}-Z_u Y_{ul})=0\,,
\end{aligned}
\end{equation}
where impedances $Z$ and admittances $Y$ are implicit functions of the mode frequency $\omega_n$. Using this dispersion relation the dependencies between transverse admittance and frequencies of modes can be investigated.
\\

With $n=0$ for the basic operational mode $\omega_0$, this equation simplifies to
\begin{equation}
\label{ZY0}
(Z_uZ_t+Z_lZ_t+Z_uZ_l)(Y_{tu}Y_{ul}+Y_{tl}Y_{ul}+Y_{tu}Y_{tl})=0\,,
\end{equation}
where one solution is $Z_uZ_t+Z_lZ_t+Z_uZ_l=0$ which defines the frequency of the $TM_{010}$-mode and of another $TM$-mode if $Z_u\not \equiv Z_l$. Equation~(\ref{ZY0}) implies that the operating frequency is independent of the transverse admittances $Y$, thus confirming that the stem configuration does not change this frequency.

The solutions with $Y_{tu}Y_{ul}+Y_{tl}Y_{ul}+Y_{tu}Y_{tl}=0$ correspond to frequencies of two different $TE_{010}$-modes, being modes where just the upper part or just the lower part of the cavity resonates. Both $TE_{010}$ modes are independent of the cavity's longitudinal tube and gap structure. For a given cavity the resonance frequencies can be determined from 3d CST~\cite{CST} simulations. Additionally, they can be expressed analytically through the dispersion curve of Equ.~(\ref{disp}), if proper fit values for $Z$ and $Y$ are plugged into the equation.
\\

\begin{figure}[hbt]
\centering
\includegraphics*[width=70mm,clip=]{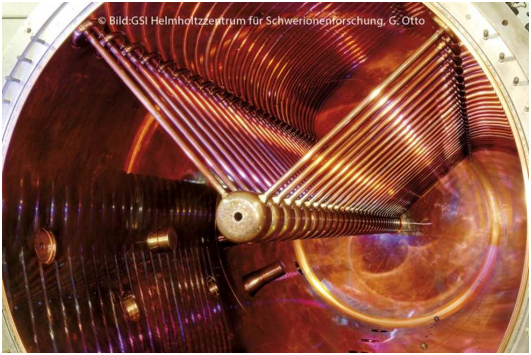}
\caption{Today's first tank of the Alvarez-type DTL of the Universal Linear Accelerator UNILAC at GSI~\cite{UNILAC}. The stems are installed in the constant V-configuration.}
\label{RealCavityf}
\end{figure}

\begin{figure}[hbt]
\centering
\includegraphics*[width=70mm,clip=]{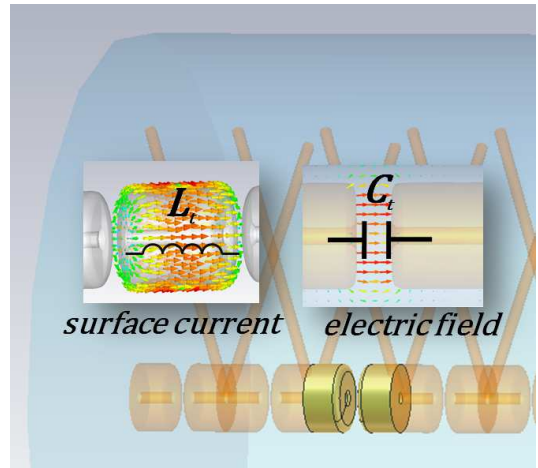}
\caption{Simulated current distribution on the drift tube surface corresponding to the inductance $L_t$ and electric gap field corresponding to the capacitance $C_t$ between tubes. The longitudinal electric field nearby the tubes is merged into this capacitance as well.}
\label{ZtLt}
\end{figure}

An Alvarez-type cavity of 8~m in length with an operating frequency $\omega_0$ of 108.4~MHz and a constant cell length of 195.6~mm was simulated. All impedances and admittances per length are constant along the beam axis. The model is based on the existing post-stripper DTL cavities of the UNILAC at GSI~\cite{UNILAC} as shown in Fig.~\ref{RealCavityf}.
The parameters for the first cavity are listed in Tab.~\ref{DTLpara}, from which we took geometric parameters for the 8 m long CST model.
\begin{table}[hbt]
\caption{\label{DTLpara} Main parameters of the first post-stripper DTL cavity of the UNILAC.}
\begin{ruledtabular}
\begin{tabular}{l|l}
Parameter & Value \\
\hline
Number of cells & 63 \\
Tube outer radius &	100 mm	\\
Tube inner radius &	15 mm\\
Cavity rf-length	& 12.4 m\\
Tank radius &	1004 mm	\\
TTF	& 0.84--0.88\\
Gap voltage	& 326--520 kV\\
Frequency & 108.4 MHz\\
Average field &	2.15 MV/m\\
E(surface) &	$\leq$ 1.0 E(Kilp.)\\
ZTT	& 32--42 MOhm/m\\
Input energy	&1.39 MeV/u\\
Output energy	&3.6 MeV/u\\
Total rf-power &	1.7 MW
\end{tabular}
\end{ruledtabular}
\end{table}
Simulations of this CST model provide rf-field maps for different modes from which explicit expressions of each impedance and admittance are extracted.
The longitudinal impedance is independent from the stem configuration. According to the tube field distribution shown in Fig.~\ref{ZtLt} the tube impedance is
\begin{equation}
Z_t= i\omega L_{t} +\frac{1}{i \omega C_t}\,,
\end{equation}
where $L_t$ is the inductance of the drift tube and $C_t$ is the capacitance of the gap. The expressions of the other elements will be established from the stem configuration of the specific case.

\subsection{Cavity without stems}
\begin{figure}[hbt]
\centering
\includegraphics*[width=85mm,clip=]{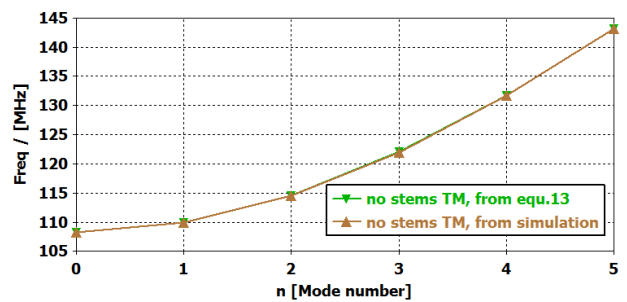}
\caption{Dispersion curves of $TM_{0n0}$-modes for an Alvarez-type cavity without stems from analytic calculation~(green) and from simulations~(brown).}
\label{NoStemDisp}
\end{figure}

If there are no stems at all, just the capacitance $C$ between tank mantle and tubes is considered in the transverse direction. The upper side and the lower side of the tank mirror each other and the expressions get
\begin{equation}
    \begin{aligned}
Z_u=Z_l= i\omega L_{z},\\
Y_{tu}=Y_{tl}=i\omega C,\\
Y_{ul}=0\,.
    \end{aligned}
\end{equation}
Equation~(\ref{disp}) then simplifies to
\begin{equation}
\label{e14}
2\frac{C}{C_t}\frac{\omega_n^2-\omega_0^2}{\omega_0^2}=\frac{n^2.\pi^2}{l^2}\,,
\end{equation}
where $\omega_n$ is the frequency of mode number~$n$.
The operating frequency ($n=0$) can be expressed analytically as
$$\omega_0\,=\,\frac{1}{\sqrt{C_t(L_t+\frac{1}{2}L_z)}}$$ from the equivalent circuit and it is known to be equal to 2$\pi\times$108.4~MHz. Equation~(\ref{e14}) fits the dispersion curve from simulations shown in Fig.~\ref{NoStemDisp} to an error of less than 0.1\%, if $C/C_t=2.57$ is assumed. Simulations and Equ.~(\ref{e14}) demonstrate that there are no $TE$-modes being exited in this case as their resonance frequencies are equal to zero.

\subsection{Cross and Alternating V-Configuration}
\begin{figure}[hbt]
\centering
\includegraphics*[width=80mm,clip=]{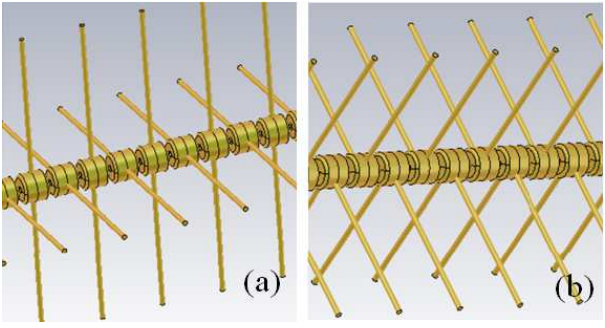}
\caption{Cross-configuration (a): the angle between two stems baring the same drift tube is 180 degree and at the subsequent tube the arrangement is rotated by 90~degrees. Alternating V-configuration~(b): the angle between two stems baring the same tube is 90~degrees and at the subsequent tube the arrangement is rotated by 180~degrees. Both stem configurations have the same stem density in the cavity quadrant resulting in equal transverse admittances. The transverse admittance between tank and tube can be reduced to zero or even to negative values.}
\label{CHstem}
\end{figure}

\begin{figure}[hbt]
\centering
\includegraphics*[width=80mm,clip=]{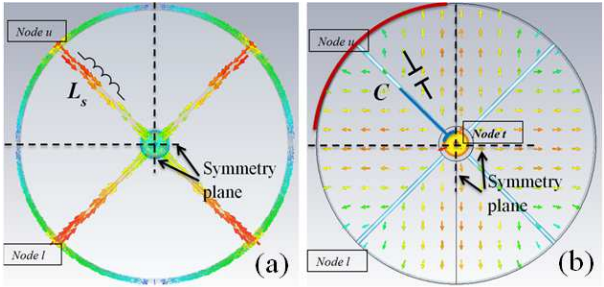}
\caption{Equivalent inductances and capacitances of one cavity cell for the cross configuration and the alternating V-configuration. (a): current path inductance $L_s$; (b): capacitance $C$ from the electric field between red cavity mantle area and blue drift tube area.}
\label{CH_LCp}
\end{figure}

In the following the cross-configuration and the alternating V-configuration as shown in Fig.~\ref{CHstem} are treated. The transverse elements of the equivalent circuit are illustrated in Fig.~\ref{CH_LCp} together with the field distribution from simulations.
\label{VTEf} 
$L_s$ is the inductance of stems and $C$ is the capacitance between the tank mantle and the tube together with some fraction of the stem. Using these equivalent elements the explicit expression for inductances and admittances are
\begin{equation}
    \begin{aligned}
Z_u=Z_l= i\omega L_{z},\\
Y_{tu}= Y_{tl}=\frac{1}{i \omega L_{s}}+i\omega C,\\
Y_{ul}= 0\,,
\label{CH_LCe}
    \end{aligned}
\end{equation}
simplifying Equ.~(\ref{disp}) to
\begin{equation}
\label{CHdispe}
\frac{C}{C_t}\frac{\omega_n^2-\omega_{e0}^2}{\omega_n^2}\frac{\omega_n^2-\omega_0^2}{\omega_0^2}=\frac{n^2.\pi^2}{l^2},
\end{equation}
There are two physical solutions for $\omega_n^2$ leading to two different dispersion curves. The first represents $TE$-modes with the basic mode frequency $\omega_{e0}$, which follows the equation $Y_{tu}(\omega_{e0})=Y_{tl}(\omega_{e0})=0$. The second dispersion function represents $TM$-modes and follows $Z_t(\omega_0)=0$.

\begin{figure}[hbt]
\centering
\includegraphics*[width=85mm,clip=]{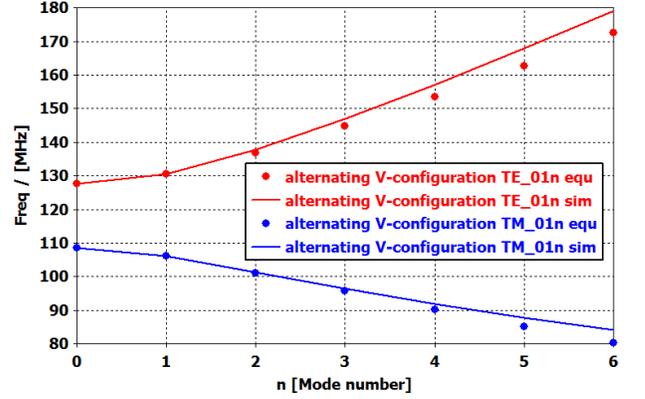}
\caption{Dispersion curves of stems with cross and alternating V-configuration. Circles are from simulations and lines are from Equ.~(\ref{CHdispe}).}
\label{CHdispf}
\end{figure}

In analogy to the procedure described in the previous subsection one obtains $\omega_{e0}= \frac{1}{{\sqrt{CL_s}}} =2\pi\times $130.0~MHz for the $TE_{010}$-mode and $\omega_{0}=2\pi\times $108.4~MHz for the $TM_{010}$-mode. Fitting the analytic dispersion function to the simulation results yields $\frac{C}{C_t}=4.061$.
The dispersion curves are plotted in Fig.~\ref{CHdispf}. The values of $L_s$ and $C$ do slightly depend on the frequency, thus causing small deviations of the simulated dispersion curves from Equ.~(\ref{CHdispe}), which neglects these dependencies.
In the cross and alternating V-configuration $\frac{C}{C_t}$ is larger than in the previous case of a cavity without stems as the capacitance $C$ between tank mantle and tubes is increased significantly by the stems. This is seen also from the field map shown in Fig.~\ref{CH_LCp}, where the part of the stem being close to the tube also provides some electric field pointing towards the tank mantle thus providing additional capacitance.

Figure~\ref{CHdispf} also shows that with this stem configuration the dispersion curve for $TM_{01(n>1)}$-bands has negative slope, the $TE$-mode frequencies are higher than the $TM$-mode frequencies, and its dispersion curve has positive slope. This is from the negative traverse admittances $Y_{tu}$ and $Y_{tl}$. The $TM_{011}$-frequency can be pushed below the $TM_{010}$-frequency. For tuning with PCs such an inversion of the dispersion curve slope happens if more PCs are applied than actually being needed for stabilization.

\subsection{Constant V-configuration}
\begin{figure}[hbt]
\centering
\includegraphics*[width=80mm,clip=]{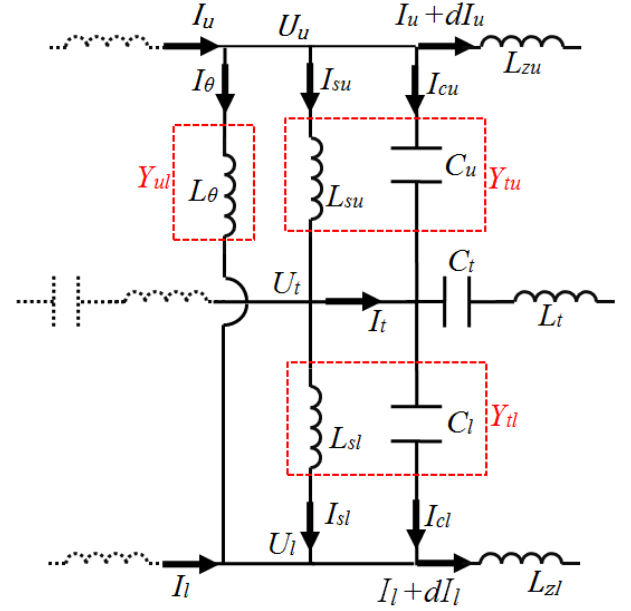}
\caption{Equivalent circuit for one Alvarez-type cavity cell with two half tubes, one gap, and one pair of stems (see also Fig.~\ref{ZtLt}). Capacitances between tank mantle and stems are merged into $C_u$ and $C_l$.}
\label{Circuit}
\end{figure}

This subsection is on the well-known case that all stems are distributed on the same side of the tank as depicted in Fig.~\ref{RealCavityf}. The transverse elements of the equivalent circuit are illustrated in Fig.~\ref{VTEf} together with the field distribution from simulations. Although a case of two stems per drift tube is shown here, the presented formalism can be applied to the single stem case as well, since the latter is the special case of an angle of zero degrees between the two stems that bare the same tube.
\begin{figure}[hbt]
\centering
\includegraphics*[width=40mm,clip=]{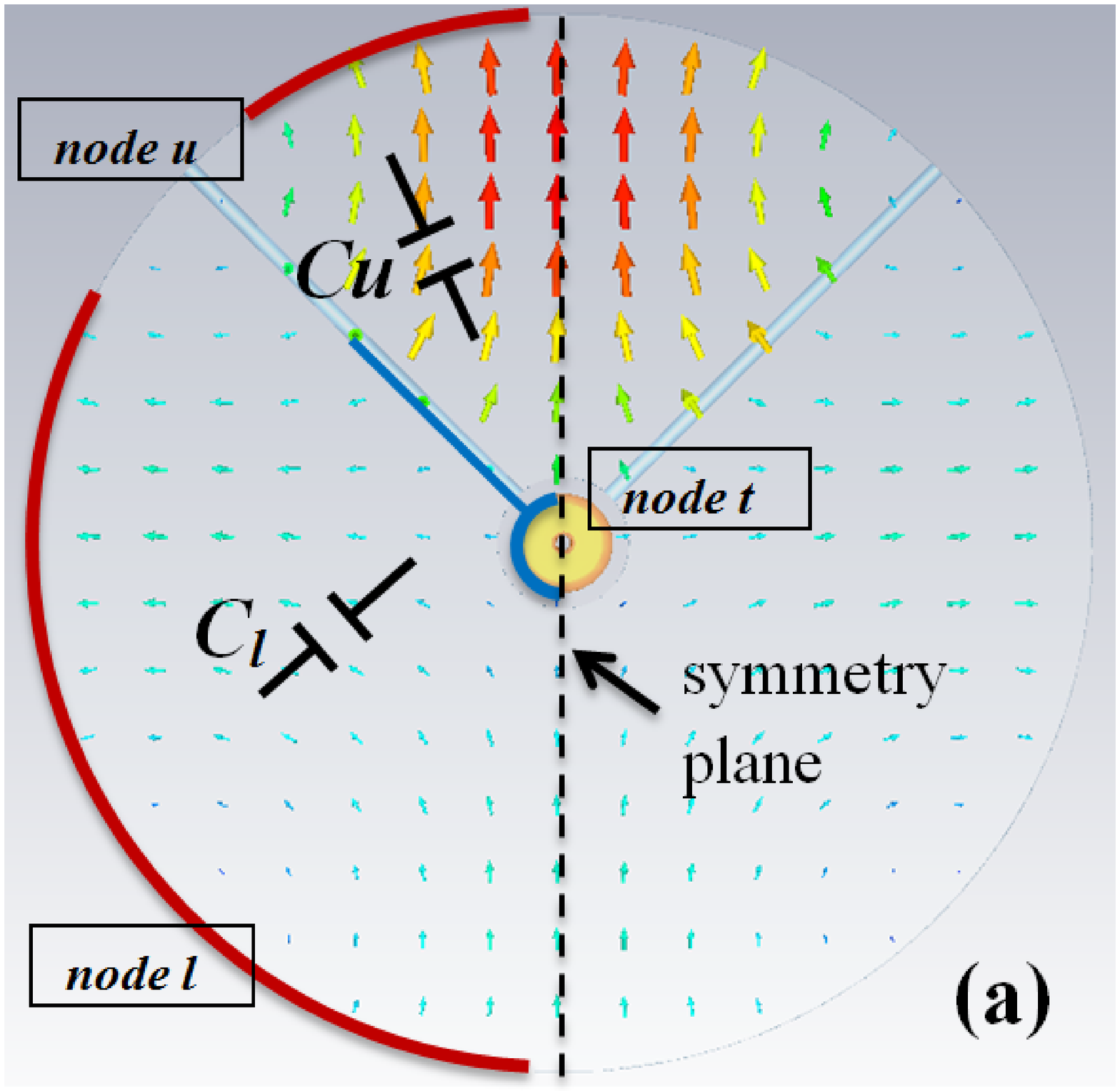}
\includegraphics*[width=39mm,clip=]{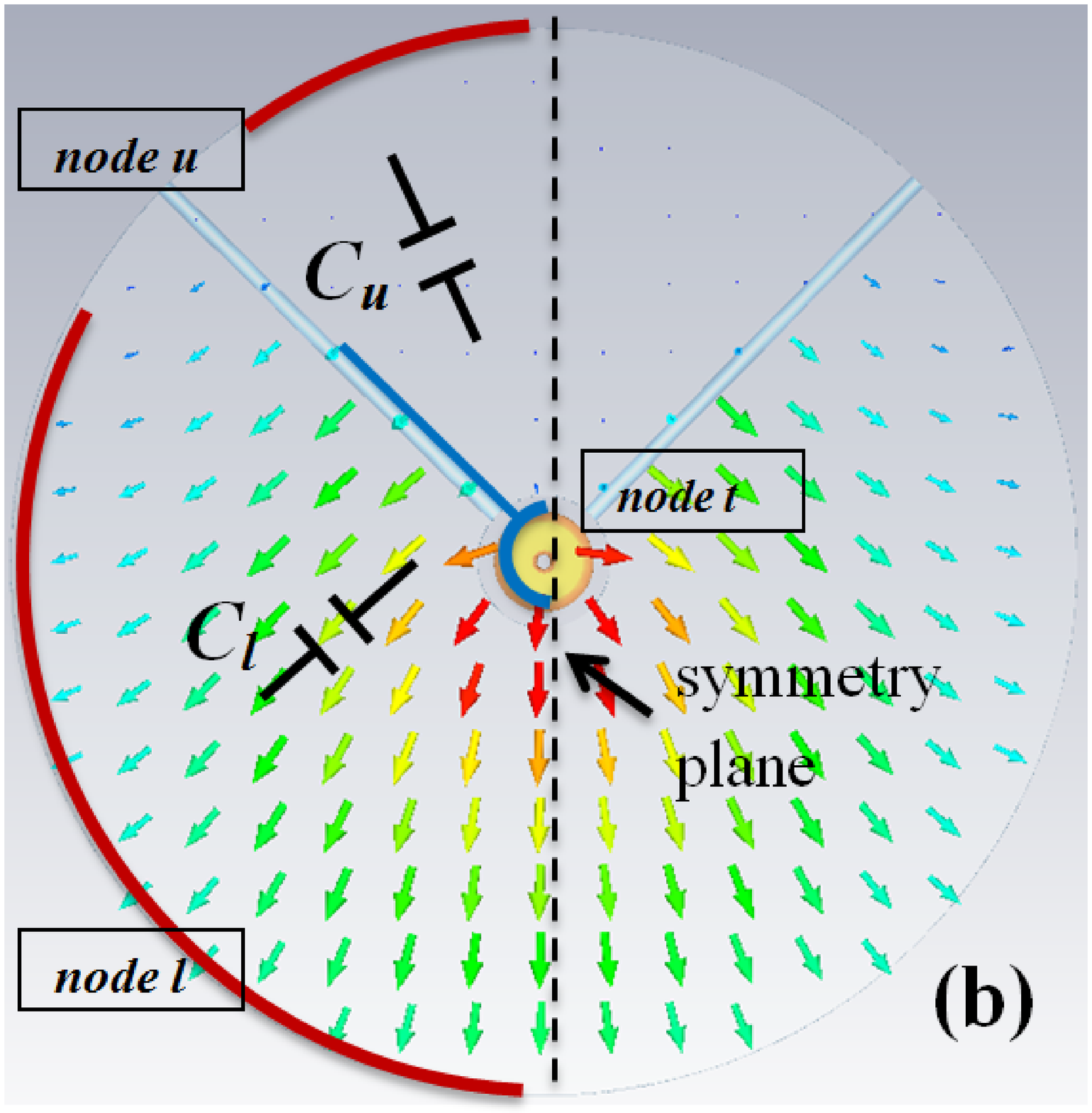}
\includegraphics*[width=40mm,clip=]{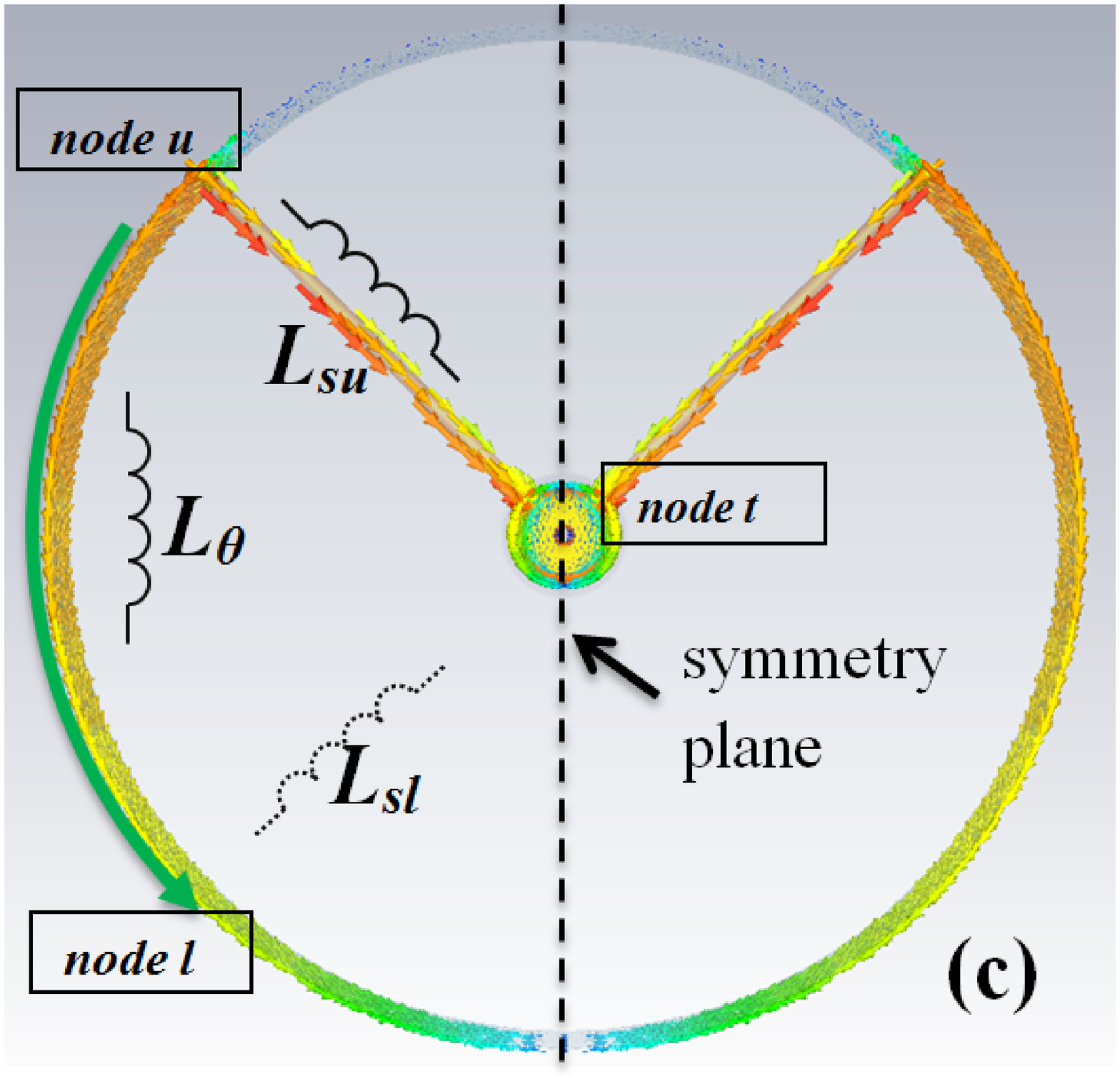}
\caption{Field and current pattern for two different $TE$-modes with constant V-configuration. (a): resonator between tube and upper part of the tank with the basic frequency of 149.2~MHz. (b): resonator between tube and lower part of the tank with the basic frequency of 65.0~MHz. (c): current pattern for case (b).}
\label{VTEf} 
\end{figure}
The equivalent circuit of an Alvarez-type cavity section with stems mounted at the upper cavity part is shown in Fig.~\ref{Circuit}.

\begin{figure}[hbt]
\centering
\includegraphics*[width=85mm,clip=]{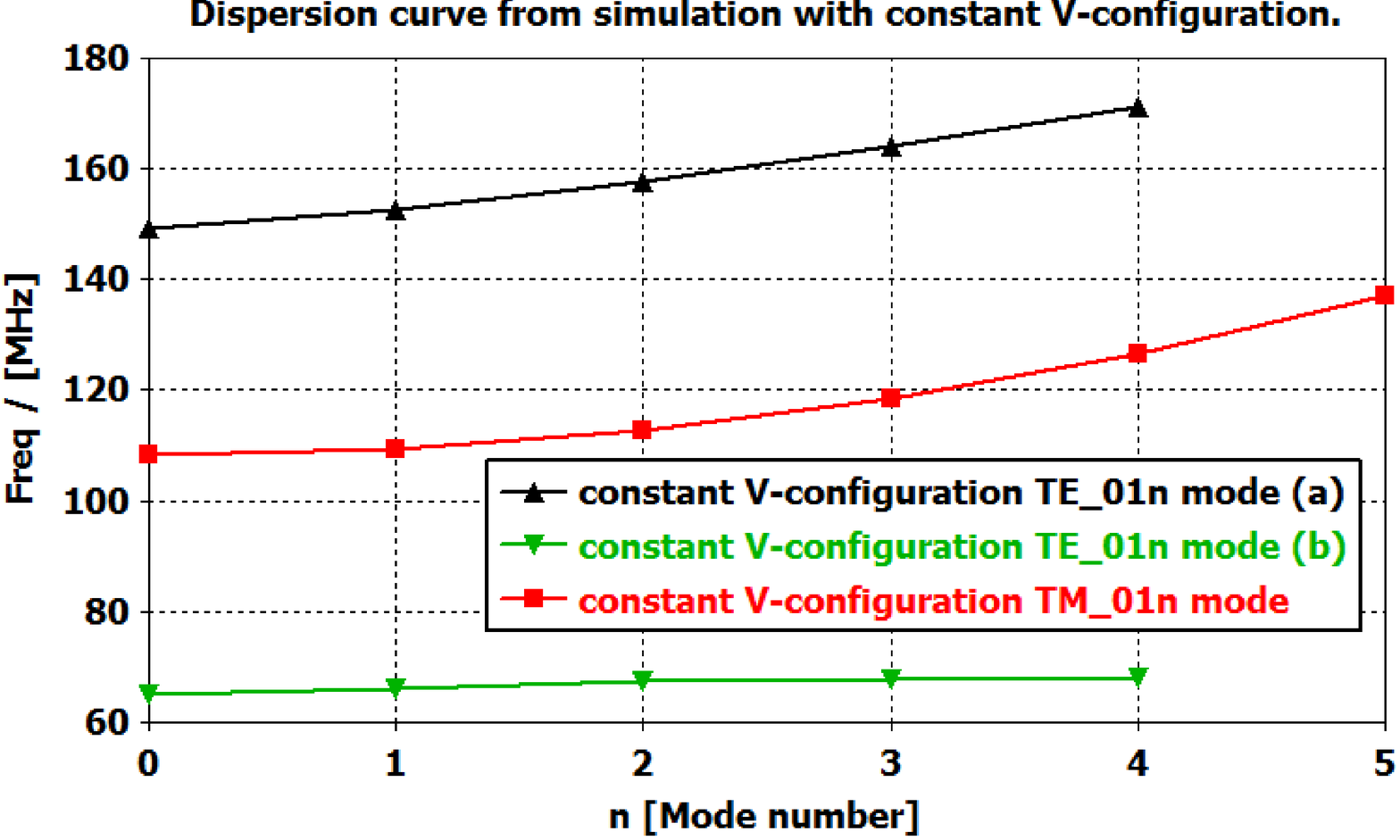}
\caption{Three dispersion curves from 3d simulation for the constant V-configuration, corresponding to three solutions of Equ.~(\ref{disp}). (a) and (b) correspond to the two modes shown in Fig.~\ref{VTEf}.}
\label{VstemDipsf}
\end{figure}

In the circuit $L_{su}$, $L_{sl}$, and $L_\theta$ are the inductances associated with the flow of longitudinal and transverse currents along the cavity wall per cell length $dz$. $C_u$ is the capacitance between the tube and the upper side of the tank, $C_l$ is the capacitance between the tube and the lower side of the tank. Both inhabit components from parts of the stems being close to the tube. $L_{su}$ and $L_{sl}$  are the inductances of the upper and lower side stems. As before in the circuit equations, constant cell length is assumed and the impedances and admittances are functions of the frequency~$\omega$. From the circuit the following expressions for longitudinal impedances per length $Z$ and transverse admittances per length $Y$ are derived:
\begin{equation}
    \begin{aligned}
Z_u= i\omega L_{zu},\\
Z_l= i\omega L_{zl},\\
Y_{tu}= \frac{1}{i \omega L_{su}}+i\omega C_u,\\
Y_{tl}= i\omega C_l,\\
Y_{ul}= \frac{1}{i \omega L_{\theta}}\,.
\label{V_LC}
    \end{aligned}
\end{equation}
Plugging these expressions into Equ.~(\ref{disp}) gives a cubic equation for $\omega_n^2$, i.e. three different solutions for each mode number $n$ lead to three different dispersion curves plotted in Fig.~\ref{VstemDipsf}. The upper dispersion curve corresponds to the resonating of the upper side of tank (Fig.~\ref{VTEf}a), and the lower curve to the resonating of the lower side of tank (Fig.~\ref{VTEf}b).
The central dispersion curve starting from 108.4~MHz represents $TM$-modes and its slope is smaller compared to the no stem case according to Equ.~(\ref{disp}).
Simulation of the constant V-configuration also revealed that if just one pair of stems is rotated by 180~degrees, the slope of the $TM_{011}$-field distribution will be inverted locally. This local inversion of $\frac{d^2U_t}{dz^2}$ is caused by the locally inverted transverse admittances. Thus one can determine the longitudinal distribution of transverse admittances by analysing the longitudinal field distribution of the $TM_{011}$-mode. This will be exploited in the tuning method introduced in the following section.

\section{Analytic field flatness stabilization}
The stem configuration determines the shapes of the branches of the dispersion curves, preserving the operational mode. According to the previous analysis, tuning through the stem configuration is by adjusting the transverse admittances to zero along the whole cavity. Tuning by post-couplers aims at closing the stop-bands, which is effectively the same thing but explained in a different way. The scaling of transverse admittances $Y_{tu}$ and $Y_{tl}$ defines the following behaviour of the cavity under tuning:
\begin{itemize}
\item field tilt sensitivity of the $TM_{010}$-mode:
if the cavity operates at nominal frequency (e.g..~108.4~MHz), the field tilt scales through the transverse admittance according to Equ.~(\ref{Etilt}).
\item slope of the dispersion curves of the $TM$-modes, notably in the interval between $TM_{010}$ and $TM_{011}$ (from Equ.~(\ref{CHdispe})): if higher $TM$-mode frequencies $\omega_{(n>0)}$ are close to the nominal mode $\omega_0$, the equation of dispersion curves simplifies to
$$
\frac{\omega_n-\omega_0}{\omega_0} \simeq \frac{\omega_0}{\omega_0-\omega_{e0}}\frac{C_t}{C}\frac{n^2.\pi^2}{2l^2}\,,
$$
implying that if $\omega_{e0}$ gets close to $\omega_0$, $\omega_{n>0}$ will be pushed away from $\omega_0$. An alternative explanation for the relationship between field tilt and frequency separation is that from perturbation theory the field distribution of the $TM_{010}$-mode with perturbations can be expressed through linear superposition of all $TM$-mode fields along a  uniform cavity
$$U_t=U_0+\sum_{n=1}^{\infty } a_{n}U_n\,,$$
where $U_n$ is the voltage distribution of $TM_{01n}$-modes without perturbation, $U_n=cos(\frac{n\pi z}{l})$, and the factor $a_n$ being proportional to $\frac{1}{\omega_n-\omega_0}$.
\item field distribution of all kinds of modes:  Equ.~(\ref{ddu}) will give a unique solution for a given transverse admittance distribution along the beam axis. For the stem configurations previously studied, the transverse admittance is constant and constant transverse admittance result in $cos(n\pi  z/l)$ like field distributions. Field distributions from admittance distributions inhabiting a slight modulation from a constant value will result in field distributions that are moderately modified but still similar to cosine like field distributions.
\item stop-band and pass-bands: lowering the transverse admittances rises the lower pass-band. At some point the stop-band disappears and the two pass-bands merge.
\end{itemize}
Two different methods will be introduced in subsection A and B to minimize the transverse admittances by well distributed stems. After tuning the field tilt sensitivity all along the cavity is strongly reduced. The tuning processes will be applied to an example cavity similar to the ones for the UNILAC upgrade considering increasing cell length.

\subsection{Tuning through $TE$-modes}
\begin{figure}[hbt]
\centering
\includegraphics*[width=80mm,clip=]{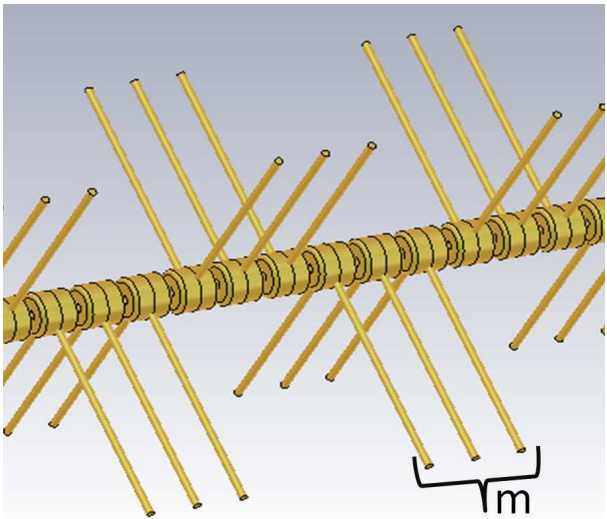}
\caption{Local minimization of transverse admittance by grouping of stems. The frequency of the $TE_{010}$-mode depends on the stem number $m$ per group. For field stabilization $m$ must be determined such that its frequency gets close to the operating frequency $\omega_0$ of 108.4~MHz.}
\label{CH_adjust}
\end{figure}

Figure~\ref{CHstem} depicts stem configurations with quadrupolar symmetry.
The capacitance $C$ between tank wall and drift tube, together with the inductance $L_s$ of the stems, comprise a resonating structure in the $TE$-mode. Its frequency $\omega_{e0}$ can be obtained from simulating a short cavity section of few cells and periodic boundary conditions. Adjusting the stem configuration of this section to $\omega_{e0} \simeq \omega_{0}$ is equivalent to tune $Y_{tu}(\omega_0)$ and $Y_{tu}(\omega_0)$ to zero along the complete cavity.
The alternating V-configuration shown in Fig.~\ref{CHstem}a features negative transverse admittances. This allows for alternating groups of stems instead of single stems, thus  adjusting locally the transverse admittances to zero. Figure~\ref{CH_adjust} shows stems being grouped and alternated accordingly. The stem number per group $m$ can be determined from simulations by requiring $\omega_{e0}\approx\omega_0$.
\begin{figure}[hbt]
\centering
\includegraphics*[width=85mm,clip=]{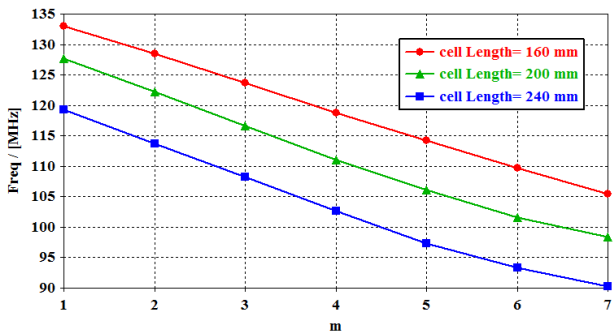}
\caption{Frequency of the $TE_{010}$-mode as a function of $m$ for different cell lengths. The parameter $m$ is to be determined such that the $TE_{010}$-mode frequency is close to the operating frequency $\omega_0$ of 108.4~MHz at each section of the cavity.}
\label{VstemTEf}
\end{figure}
To account for increasing cell length, the stem configuration needs adoption as $\beta\lambda$ increases. Figure~\ref{VstemTEf} shows the frequency of the $TE_{010}$-mode as a function of $m$ for different $\beta\lambda$. $m$ is to be chosen such that the frequency of the $TE_{010}$-mode becomes equal to the nominal operating frequency, being 108.4~MHz in the actual example. The procedure to obtain the optimum stem configuration is:
\begin{description}
\item[1)] a very short cavity of few cells with constant V-configuration is modelled. The stems at the tank entrance are rotated by 180~degrees towards the bottom half of the tank. Thus the stems are grouped into two sections, with stem number $m$ in each group. Periodic boundary conditions are defined at the end walls. The cell length and $m$ are used as the adjustable parameter.
\item[2)] $\beta\lambda$ is fixed for the complete cavity and the frequency of the basic $TE$-mode is determined from simulations. If it is lower (higher) than the operational mode, the number of cells is reduced (increased) and step~1) is repeated until the frequency of the basic $TE$-mode comes as close as possible to the operating frequency~$\omega_0$.
\item[3)] After finishing the design of the first cavity section, the cell length of the subsequent section is fixed and the procedure of step~2) is applied again.
\item[4)] After finishing the last section, simulation of the complete cavity with conducting wall boundary is performed. Frequencies of modes close to the operational mode are checked as well as the field distribution of the $TM_{011}$-mode.
\end{description}

\begin{figure}[hbt]
\centering
\includegraphics*[width=80mm,clip=]{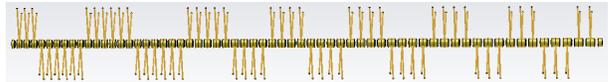}
\caption{Stem configuration {\bf E} determined through tuning the frequency of the basic $TE$-mode. The increase of the cell length leads to a decrease of the stem density and $m$ decreases accordingly.}
\label{StemDes2}
\end{figure}

The method takes advantage of short cavity simulations with high accuracy instead of many simulations of the complete cavity with reduced accuracy.
Figure~\ref{StemDes2} depicts the final stem configuration, referred to as {\bf E} in the following. The frequency of the $TM_{011}$-mode is shifted from 108.4+0.5~MHz to 108.4+4.1~MHz.

\subsection{Tuning through $TM_{011}$-mode}
\begin{figure}[h]
\centering
\includegraphics*[width=85mm,clip=]{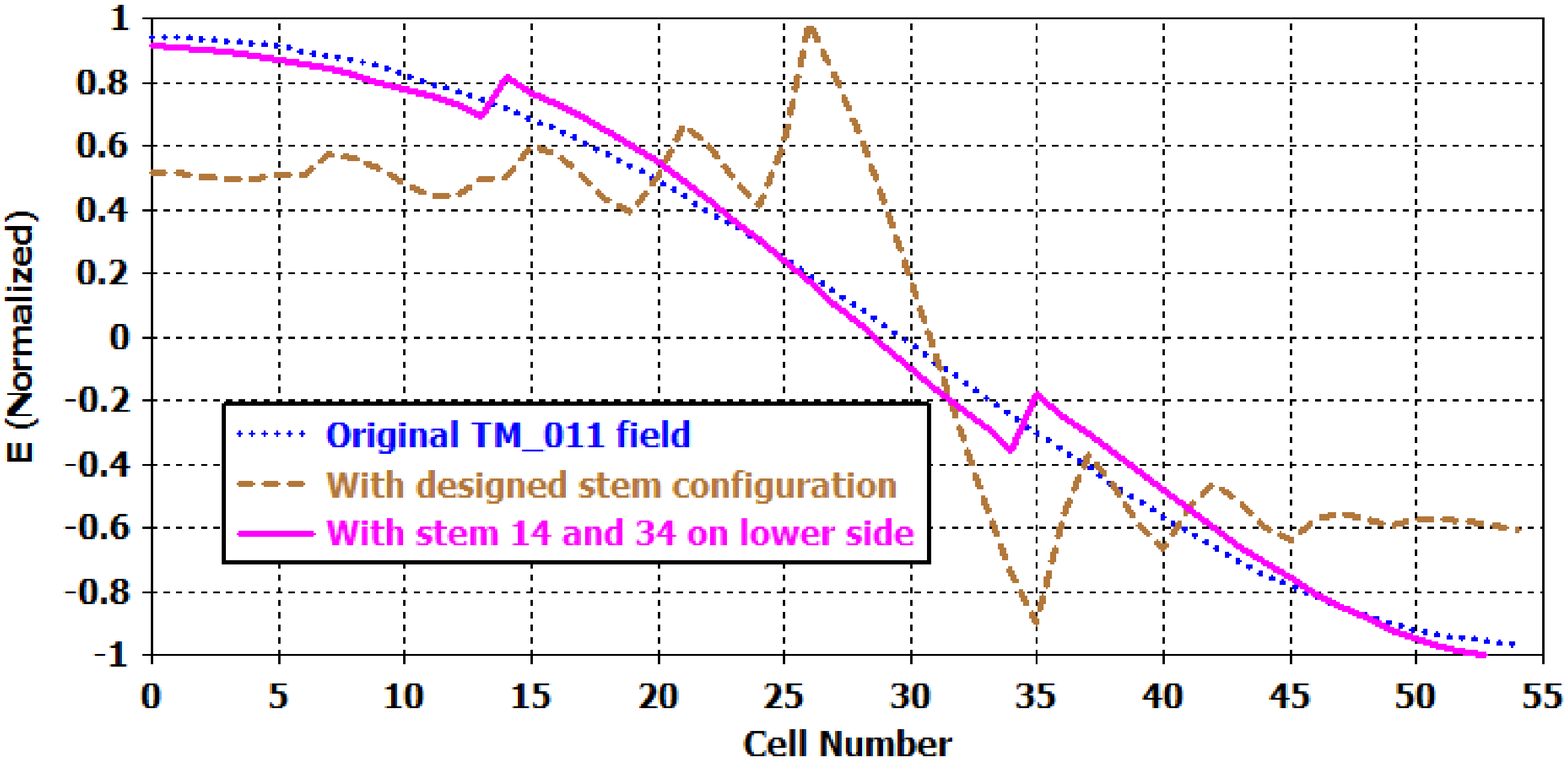}
\caption{Longitudinal electrical field strength of the $TM_{011}$-mode as a function of the cell number for an Alvarez-type cavity for three different cases: original constant V-configuration of stems (blue), original constant V-configuration with the stem pairs of cell~14 and~34 being rotated by 180~degrees (purple), and final stem configuration after optimisation w.r.t. local field flatness as shown in Fig.~\ref{StemDes1}.}
\label{TM011Field}
\end{figure}

\begin{figure}[h]
\centering
\includegraphics*[width=85mm,clip=]{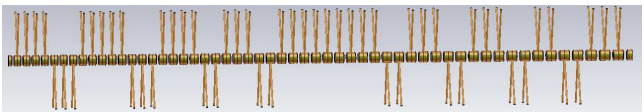}
\caption{Stem configuration {\bf M} determined through tuning the local flatness of the $TM_{011}$-mode.}
\label{StemDes1}
\end{figure}

As mentioned in the previous section, for the constant V-configuration rotation of any pair of stems by 180~degrees will locally change the sign of the slope of the field distribution of the $TM_{011}$-mode as shown in Fig.~\ref{TM011Field}. It is from the local change of sign of the expressions for $Y_{tu}$ and $Y_{tl}$ in Equ.~(\ref{Etilt}). As the frequency of the $TM_{011}$-mode is very close to $\omega_0$, it can be rewritten as $\omega_1=\omega_0+\delta \omega$. If at some location $z$ along the axis $Y(\omega_0+\delta \omega)=0$, the field tilt of the $TM_{011}$-mode becomes equal to zero. As $Y(\omega_0)\simeq 0$ is a reasonable approximation, the field tilt of the $TM_{010}$-mode on that location will be cancelled as well. For the ideal model the field distribution of the $TM_{011}$-mode is $E\propto cos(\pi z/l)$. The proposed tuning method shall modify this cosine function to a function being closer to a step function, preserving the initial and final value of the original cosine function. The smooth cosine like inversion of the initial field strength at the cavity entrance to its negative value at the cavity exit will be re-shaped towards a more abrupt change of polarity close to the center of the cavity. Re-shaping is done by profiting from the fact that a 180~degree rotation of the stems changes the sign of the local slope of the $TM_{011}$-field distribution. Proper choice of the local density of rotated stems will thus locally flatten the field distribution. The procedure to obtain the optimum stem configuration is:
\begin{description}
\item[1)] Simulations of a complete cavity with constant V-configuration are performed to obtain the field distribution of the $TM_{011}$-mode.
\item[2)] A group of stems located not far from each end wall is rotated by 180~degrees towards the bottom of the tank. With simulations it is checked whether the bottom $TM_{011}$ field distribution is locally flattened. If not, the number of stems forming the group and/or its position is varied.
This step is done from both sides of the cavity simultaneously.

\item[3)] Step~2) is repeated with the next two groups towards the centre of the cavity.
\item[4)] Step~3) is repeated until two flattened sections of the field connected by a step close to the cavity centre are produced as shown in Fig.~\ref{TM011Field}.
\end{description}
Figure~\ref{TM011Field} shows the $TM_{011}$-field distribution and how it is changed by modifying the constant V-configuration of the stems.
%
The final stem configuration, referred to as {\bf M} in the following, obtained by applying this method is depicted in Fig.~\ref{StemDes1}.
For this stepwise flattened field distribution the frequency of the $TM_{011}$-mode is shifted from 108.4+0.5~MHz to 108.4+4.5~MHz. The effect of the proposed tuning procedures on the operational mode is evaluated in the following section.

\subsection{Verification through Simulations}
\begin{figure}[h]
\centering
\includegraphics[width=85mm,clip=]{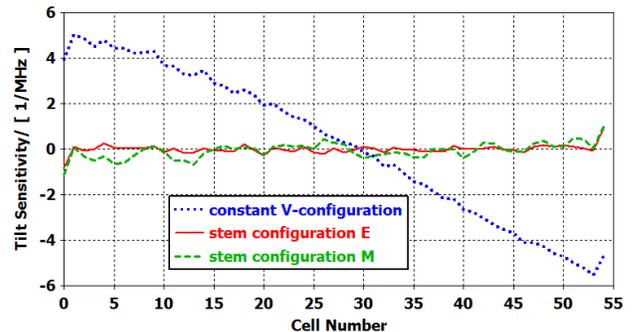}
\caption{Field tilt sensitivity of the operational mode as defined in Equ.~(\ref{tilt_sensitivity}) of an Alvarez-type cavity for different stem configurations. The applied perturbations are described in the text. Blue: a constant V-configuration without stabilization through stem rotation. Red: stabilization through stem configuration {\bf E} of Fig.~\ref{StemDes2}. Green: stabilization through stem configuration {\bf M} of Fig.~\ref{StemDes1}.}
\label{PerturbationCheck}
\end{figure}

To evaluate the efficiency of the two tuning methods, a perturbation is applied to cavities featuring designs {\bf E} and {\bf M}, respectively. The length of the first gap is reduced by 2~mm to 25.5~mm, causing a frequency shift $\Delta f$ of -140~kHz. Then the length of the last gap is increased by 4~mm to 53~mm such that the total frequency shift is zero. The TS of cell $i$ is defined as
\begin{equation}
\label{tilt_sensitivity}
TS = (E^{pert}_{0i}-E^{unpert}_{0i})/(E^{unpert}_{0i}\cdot\Delta f)\,,
\end{equation}
where $E_{0i}$ is the average field strength of cell $i$. Figure~\ref{PerturbationCheck} plots the TS for different stem configurations. Both stem configurations {\bf E} and {\bf M} reduce significantly, i.e. they practically remove, the TS compared to the original constant V-configuration. Simulations showed that even for the extreme case that the first drift tube is removed from the cavity, the field distortion is kept below $\pm 10\%$. A constant V-configuration without stabilization shows a field distortion of more than $\pm 65\%$ in that case.
The two field-stabilizing stem configurations obtained from two tuning strategies result into a flat field distribution being very resistant to perturbations of the cavities geometry.

\begin{figure}[h]
\centering
\includegraphics*[width=85mm,clip=]{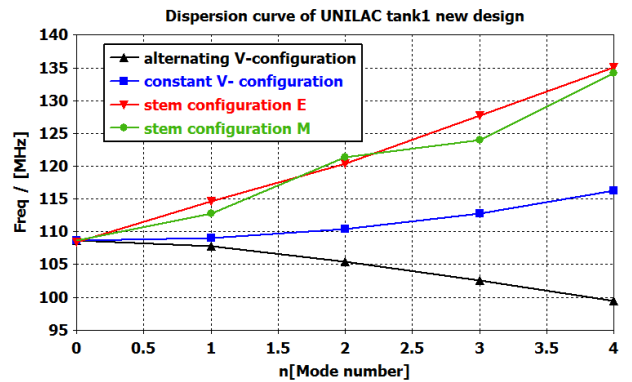}
\caption{Simulated dispersion curves of an Alvarez-type cavity with different stem configurations. Blue: constant V-configuration without stabilization. Red: stem configuration {\bf E} of Fig.~\ref{StemDes2}. Green: stem configuration {\bf  M } of Fig.~\ref{StemDes1}. Black: alternating V-configuration.}
\label{tank1dispf}
\end{figure}

The optimized stem configurations modify the dispersion curves of the $TE$-modes by increasing their slopes as summarized in Fig.~\ref{tank1dispf}. The nearest mode is shifted away from the operational mode.

\section{Conclusion}
Tuning of Alvarez-type cavities w.r.t. flat longitudinal field distribution with very low field tilt sensitivity was achieved in simulations. The method does not use any post-couplers which impose additional cost and rf-power losses. The required optimizations of the cavity rf-parameters were accomplished by adjusting the azimuthal orientations of the stems that bare the drift tubes. Two methods were developed: division of the cavity into short subsections and adjusting their local $TE_{010}$-mode frequencies to the operational frequency. Usage of short sections significantly reduces the time for simulations. The second method transforms the longitudinal field dependency along the cavity of the $TM_{011}$-mode from a cosine like half wave to a step like function. Both methods effectively minimize the transverse admittances and maximize the slope of the $TM$-modes dispersion curve, which is understood from analysis of the cavity equivalent circuit including stems and drift tubes. Both schemes lead to stem configurations that provide very low field sensitivity. The validity of the methods was benchmarked through extensive simulations.

\section{Acknowledgement}
The authors wish to express their gratitude to S.~Ramberger for several fruitful discussions and advices.

\bibliography{C.Xiao}
\end{document}